\documentclass[11pt,notoc]{JHEP3}
\usepackage{latexsym}
\usepackage{graphicx}
\usepackage{amsmath,amssymb}

\def\be{\begin{equation}}
\def\ee{\end{equation}}
\newcommand{\bea}{\begin{eqnarray}}
\newcommand{\eea}{\end{eqnarray}}

\newcommand{\orin}{\Omega_r^{(\text{in})}}
\newcommand{\omin}{\Omega_m^{(\text{in})}}
\newcommand{\orout}{\Omega_r^{(\text{out})}}
\newcommand{\omout}{\Omega_m^{(\text{out})}}

\renewcommand{\in}{{(\text{in})}}
\newcommand{\out}{{(\text{out})}}
\newcommand{\ain}{{a^{(\text{in})}}}
\newcommand{\aout}{{a^{(\text{out})}}}
\newcommand{\dec}{*}

\title{Do primordial lithium abundances imply there's no Dark Energy?
}

\author{Marco Regis$^{1,2}$ and Chris Clarkson$^1$\\
$^1$Astrophysics, Cosmology \& Gravitation Centre, and, Department of Mathematics \& Applied Mathematics, University of Cape Town, Rondebosch 7701, Cape Town, South Africa\\ $^2$Centre for High Performance Computing, 15 Lower Hope St, Rosebank, Cape Town, South Africa\\
E-Mail: \email{marco.regis@uct.ac.za,chris.clarkson@uct.ac.za}}

\abstract{
Explaining the well established observation that the expansion rate of the universe is apparently accelerating is one of the defining scientific problems of our age. Within the standard model of cosmology, the repulsive `dark energy' supposedly responsible has no explanation at a fundamental level, despite many varied attempts. A further important dilemma in the standard model is the lithium problem, which is the substantial mismatch between the theoretical prediction for $^7$Li from Big Bang Nucleosynthesis and the value that we observe today. This observation is one of the very few we have from along our past worldline as opposed to our past lightcone. By releasing the untested assumption that the universe is homogeneous on very large scales, both apparent acceleration and the lithium problem can be easily accounted for as different aspects of cosmic inhomogeneity, without causing problems for other cosmological phenomena such as the cosmic microwave background. We illustrate this in the context of a void model.
}


\begin{document}

\section{Introduction:}
Distant supernovae Ia (SNIa) explosions are too dim to be explained when taken with other cosmological observations, a fact based on a model where the curvature of the universe is constant in space and all material in the universe is gravitationally attractive.  The standard interpretation of SNIa observations within the concordance model of cosmology is that the cosmological constant, $\Lambda$, takes on about the same value as the total present day energy densities of cold dark matter (CDM) and baryons in the Universe. This extraordinary coincidence of numbers, at odds with quantum field theory estimates for the vacuum energy by 120 orders of magnitude, leaves many cosmologists slightly queasy. Attempts to explain such values typically postulate the physical existence of an infinite number of universes with randomly chosen constants of nature, whereby it just so happens that we live in a universe like ours at the particular time when $\Lambda$ starts to dominate. An elegant idea, but untestable except in highly specialised configurations. 
Alternative explanations for the observed brightnesses of SNIa within the homogeneous Friedmann-Lema\^\i tre-Robertson-Walker (FLRW) models ether postulate exotic forms of matter with huge negative pressures at low temperature, or make ad hoc changes to Einstein's General Relativity on Hubble scales~\cite{caldwell}.

A seemingly radical explanation for the SNIa observations suggests that the universe is not exactly homogeneous on Hubble scales, but that instead there is significant spatial variation in the matter distribution, which is accompanied by spatial changes in the curvature and local expansion rate. In the simplest of these models, we live in a region where the matter density is significantly less than the density of that of the universe on super-Hubble scales~\cite{Moffat:1994qy}. The simplicity of the models means that they require violation of the Copernican Principle, that we are in some sense `typical' observers. Consequently, they suffer a spatial coincidence problem rather than the temporal one of concordance cosmology. Nevertheless, these models fit SNIa and other local background observations~\cite{celerier,gbh1,February:2009pv}, and, as we show, the cosmic microwave background (CMB). Structure formation remains the unexplored area where these models may fail standard tests, but this is a technically challenging problem which has not been properly attempted. The early universe is a key area to be analysed, but this is often assumed trivial because the models can evolve from FLRW. It's not. \footnote{These models are often mistakenly dismissed because of a so-called weak singularity the models exhibit when made to reproduce a negative deceleration parameter at the origin~\cite{VFW}. It is not relevant for cosmological modelling because we don't measure a negative deceleration parameter directly, we infer it from a parameterised FLRW model. It doesn't occur if we fit a void model to data directly.}

\section{Big-Bang nucleosynthesis}
Big-Bang nucleosynthesis (BBN) is the most robust probe of the first instants of the post-inflationary Universe. After three minutes, the lightest nuclei (mainly D, $^3$He, $^4$He, and $^7$Li) were synthesised in observationally significant abundances~\cite{Steigman:2007xt,Iocco:2008va}. Observations of these abundances provide powerful constraints on the primordial baryon-to-photon ratio $\eta=n_b/n_\gamma$, which is constant in time during adiabatic expansion. In the $\Lambda$CDM model, the CMB constrains $\eta_{CMB}=6.226\pm0.17\times10^{-10}$~\cite{Iocco:2008va} at a redshift $z\sim1100$. Observations of high redshift low metallicity quasar absorbers tells us D/H$=(2.8\pm0.2)\times 10^{-5}$~\cite{Pettini:2008mq} at $z\sim3$, which in standard BBN leads to $\eta_D=(5.8\pm0.3)\times10^{-10}$, in good agreement with the CMB constraint. 
In contrast to these distant measurements at $z\sim 10^3$ and $z\sim3$, primordial abundances at $z=0$ are either very uncertain (D and $^3$He), not a very sensitive baryometer ($^4$He), or, most importantly, in significant disagreement with these measurements~-- $^7$Li.
To probe the BBN yield of $^7$Li, observations have concentrated on old metal-poor stars in the Galactic halo or in Galactic globular clusters.
The ratio between $\eta_{Li}$ derived from $^7$Li at $z=0$ and $\eta_D$ derived from $D$ at $z\sim3$ is found to be $\eta_D/\eta_{Li}\sim 1.5$.  Within the standard model of cosmology, this anomalously low value for $\eta_{Li}$ disagrees with the CMB derived value by up to 5-$\sigma$~\cite{Cyburt:2008kw}.

Several attempts at explaining this discrepancy have been done relying on systematic errors in the derived abundance
(mostly related to parameters of the stellar atmospheres such as, e.g., the temperature scale), and uncertainty in the estimates of nuclear reaction rates.
On the other hand, such corrections are typically found to be much smaller than the gap between predictions in standard cosmology and observations (see, e.g., Refs.~\cite{asplund} and \cite{Cyburt:2008kw}).
Another hypothesis is that $^7$Li has been depleted even in metal-poor stars. A strong initial astration (i.e., ``uniform" depletion affecting the whole Milky Way or even the whole Universe) appears rather unlikely (e.g.,~\cite{spite}). Depletion in the stellar atmospheres seems instead more feasible. In the globular cluster NGC 6397~\cite{korn} and in a sample of 88 metal poor stars~\cite{melendez}, it has been found that describing the depletion through atomic diffusion processes partially compensated by turbulence~\cite{richard} can lead to a $^7$Li abundance in concordance with cosmological predictions. However, further confirmations are required since this mechanism largely depends on a few fine tuned free parameters (see, e.g., the discussion in \cite{sbordone} and \cite{monaco}), depletion by diffusion/turbulence in globular clusters has been questioned \cite{hernandez}, and the discrepancy still remains in field metal-poor dwarfs when the same ``depletion correction" of Ref.~\cite{korn} is applied~\cite{spite}.
Therefore, the state of the art in the field is that there is no widely accepted robust astrophysical solution to the $^7$Li problem.

\begin{figure*}[t]
   \centering
   \includegraphics[width=0.9\textwidth]{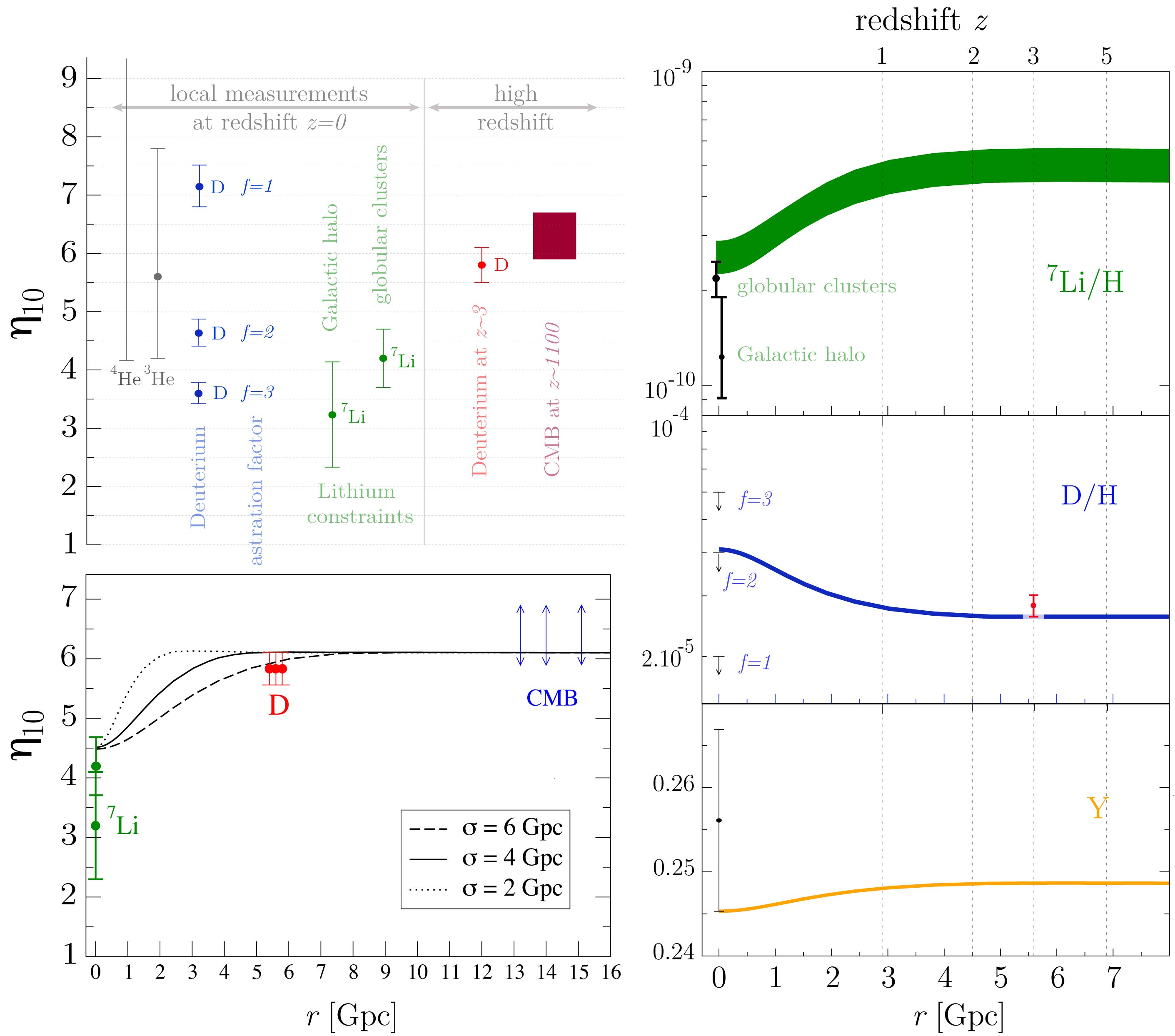}
\caption{{\bf Constraints on $\eta$.} Top left we estimate current constraints on $\eta_{10}=10^{10}\eta$ from different observations. Constraints from $^7$Li observations~\cite{Cyburt:2008kw} in Galactic globular clusters and Galactic halo are shown separately, alongside $^4$He~\cite{aver} and $^3$He~\cite{Steigman:2007xt}. These agree with each other if $\eta_{10}\sim 4.5$. Local measurements of D show a significant scatter~\cite{Iocco:2008va} though they also require $\eta_{10}\sim4.5$ if we consider the value from Bayesian analyses in~\cite{Prodanovic:2009es} and an astration factor of $f\sim 2$.
On the other hand, D observations at high redshift (red)~\cite{Pettini:2008mq} and CMB require $\eta_{10}\simeq 6$. Bottom left we show how a varying radial profile for $\eta_{10}$ (from $\sim4.5$ at the center to $\sim6$ asymptotically) can fit all the observational constraints, for differing inhomogeneity scales. (The D and CMB constraints are in redshift, so move when given in terms of comoving distance $r$, since $r(z)$ is dependent on inhomogeneity profile.) On the right we show the nuclei abundances as a function of $z$ in our example model (see text). Filling in points on this graph will test this theory.}
\label{fig:eta}
\end{figure*}

On the other hand, a local value of $\eta\sim 4-5\times10^{-10}$ is consistent with all the measurements of primordial abundances at $z=0$ (see top left panel in Fig.~\ref{fig:eta}). The disagreement with high-redshift CMB and D data (probing $\eta$ at large distances) shows up only when $\eta$ is assumed to be homogeneous on super-Hubble scales, as in standard cosmology. An inhomogeneous radial profile for $\eta$ can thus solve the $^7$Li problem, as we show in Fig.~\ref{fig:eta}.

The local primordial abundances of $^7$Li~\cite{Cyburt:2008kw}, $^4$He~\cite{aver}\footnote{For the $^4$He mass fraction, we consider $Y=0.2561\pm0.0108$, recently derived in~\cite{aver}, as reference value. Previous measurements pointed towards lower values, e.g., $Y=0.240\pm0.006$ reported in \cite{Steigman:2007xt} which is compatible as well with our benchmark scenario. On the other hand, another recent analysis \cite{Izotov:2010ca} found $Y=0.2565\pm0.001\pm0.005$, namely, a central value similar to \cite{aver}, but with significantly smaller error bars. This would be in tension with our model as well as with the standard cosmological model, and we consider $Y$ from \cite{aver} as a more conservative estimate.} and $^3$He~\cite{Steigman:2007xt} agree with each other if $\eta_{10}\sim 4.5$ (which we consider as our benchmark value for local $\eta$ in the rest of the paper).
In this scenario, local measurements of D in the interstellar medium (ISM)~\cite{Prodanovic:2009es} require in turn an astration factor (i.e. depletion of D due to star formation in our Galaxy) of $f\simeq 2$.
This value lies slightly above the range $1.4\leq f\leq1.8$ predicted in the simulation of~\cite{Romano:2006uy}. However, given the 
number of assumptions involved in a theoretical model of Galactic chemical evolution (see, e.g., the discussion in~\cite{Romano:2009cq,sembach} and reference therein), there is no strong argument to exclude $f\sim2$ (conservatively, the range $f=1-3$ can satisfy current constraints; as an example see~\cite{fields}). Moreover, observations of D abundances in the ISM show a significant scatter~\cite{Iocco:2008va}; for sake of concreteness, we reported the value from the analysis in Ref.~\cite{Prodanovic:2009es}, from which the astration factor follows. Considering a slightly higher estimate for the mean local value of D/H leads to smaller $f$ (cf, Fig.~\ref{fig:eta}), while remaining consistent with $^7$Li observations.

A Gpc-scale inhomogeneity in the baryon-to-photon ratio can therefore explain the mismatch between $\eta_{Li}$ and $\eta_{CMB}$, without violating other constraints on nuclei abundances. It should be possible to construct a model of such a variation of $\eta$ within the standard model using an $O(1)$ isocurvature mode at early times. However, the fact that such a mode must be effectively localised about us (if it wasn't homogeneous at CMB distances it would leave a significant distortion in the CMB), and would generate curvature perturbations at late times when it enters the Hubble radius (curvature and isocurvature modes are only independent on super-Hubble scales), suggests a deeper connection with models that can explain supernovae distances without dark energy.

\section{Void models with inhomogeneous $\eta$}
How could dark energy be related to the lithium problem? Let us consider this in the context of a toy void model which can fit SNIa data, though our discussion is more widely applicable. Spherically symmetric void models have a spatially varying curvature, with differing values today at the centre and at large distances (where we observe the CMB). The scale of the void is of order the Hubble radius today, so is only directly significant at late times. When the curvature mode is larger than the horizon size, gradients in the curvature across a distance $H(a)^{-1}$ are negligible; significant effects appear between widely separated radii. For the BBN and local physics at decoupling in CMB calculations we assume that early time evolution in a causal patch follows the usual FLRW evolution, with a different FLRW model at each radius.

\begin{figure*}[t]
   \centering
   \includegraphics[width=0.9\textwidth]{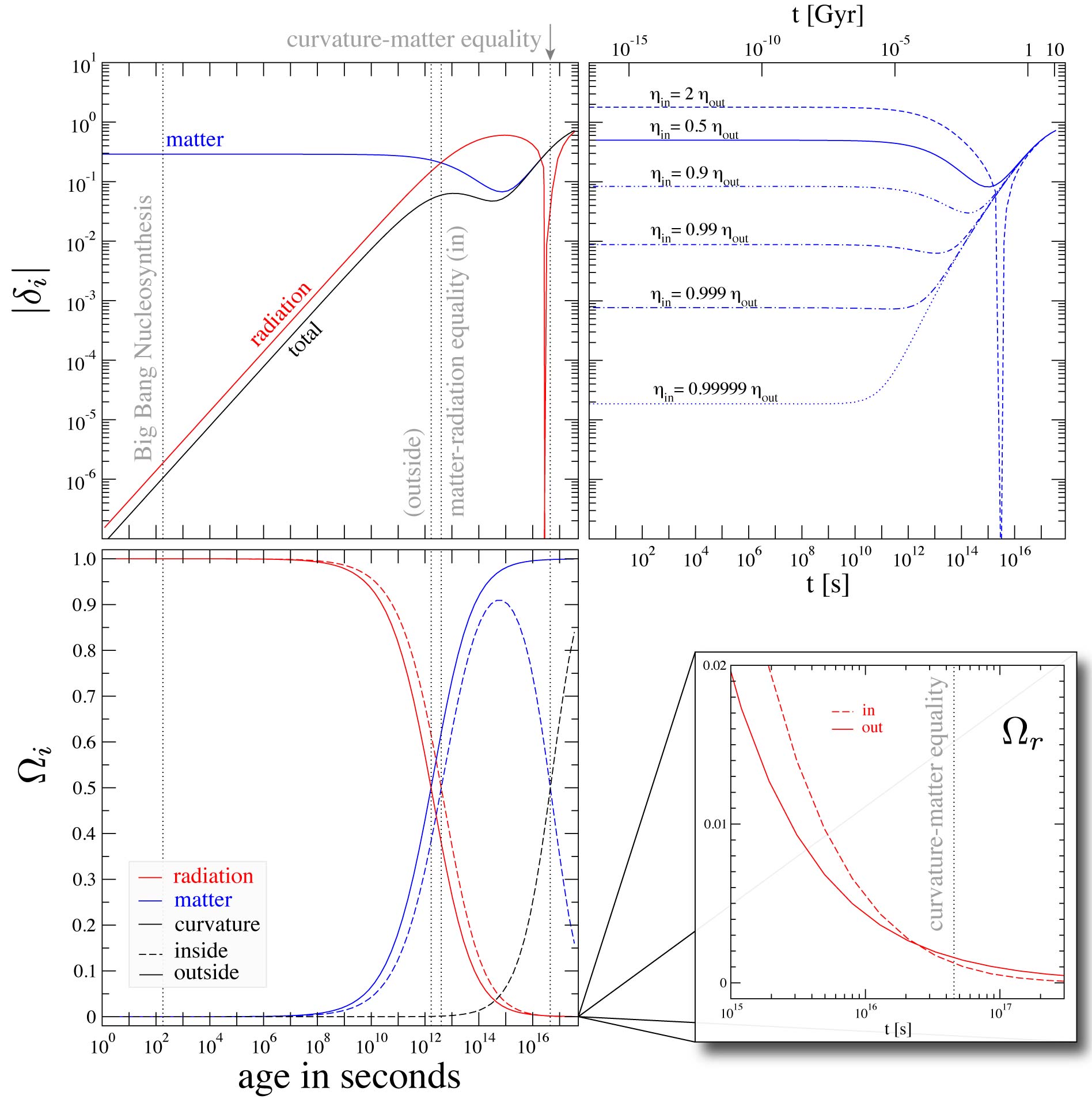}
\caption{{\bf The evolution of inhomogeneity}, for an asymptotically flat version of the example model.
If we evolve an inhomogeneous model back to BBN time any inhomogeneity $\delta_r=1-\rho^\in/\rho^\out$ in the radiation component today decays (top left), so that the radiation is homogeneous at early times during the radiation era. Meanwhile, the inhomogeneity $\delta_m$ present in the matter content decays as we head back in time to matter-radiation equality and then grows and freezes. This freezing  accounts for the $^7$Li abundances. On the upper right we show the effect of changing $\delta\eta$: if this is small the early time matter inhomogeneity $\delta_m$ follows the radiation.
On the bottom left, we show the evolution of $\Omega_i^\in$ and $\Omega_i^\out$. The zoomed plot shows the radiation profile changing from an overdensity to an underdensity at late times. The evolutions of $\rho^\in$ and $\rho^\out$ follow from standard FLRW evolutions since spatial gradients are negligible at the center and at large distances. Corrections to this assumption would only show up in the dashed line in the blow-up, after curvature matter equality.}
\label{fig:delta_evol}
 \end{figure*}

 At early times during the radiation era, we find that unless $\eta$ is fine-tuned to be exactly spatially constant, the radiation inhomogeneity grows while the matter inhomogeneity is frozen at a value $\delta_m\simeq1-\eta^\in/\eta^\out\gg\delta_r$ (where $\in$ refers to quantities at the center and $\out$ at asymptotic distances, and $\delta_i=1-\rho_i^\in/\rho_i^\out$ with $i=$matter or radiation, and we assume for simplicity that the baryon fraction $f_b$ is spatially constant). This is shown in Fig.~\ref{fig:delta_evol} and is the key to linking the lithium and dark energy problems.

It is not difficult to derive explicitly the early time evolution of the inhomogeneity.
At early times curvature is negligible and $\rho_{\text{m}}\ll\rho_{\text{r}}$. One can evaluate $\delta_i$ at fixed cosmic time by comparing $H^\in/H^\out=d\ln a^\in/d\ln a^\out$ which, neglecting spatial gradients, leads to
\bea
\frac{da^\in}{da^\out}&\simeq&\left(\frac{T^\in_0}{T_0^\out}\right)^{2}\frac{\aout}{\ain}\\
&\times&\left[1+\frac{1}{2}\left(\frac{\omin}{\orin}\right)\ain-\frac{1}{2}\left(\frac{\omout}{\orout}\right)\aout \right] \nonumber,
\eea
to first-order in $a\Omega_m/\Omega_r=\rho_m/\rho_r$,  where the $\Omega$'s always represent todays values.  Then, integrating and dropping a decaying mode,
\be
\frac{\ain}{\aout }\simeq \left(\frac{T^\in_0}{T^\out_0}\right)\left[1+\frac{1}{3}\aout
\left(\frac{T^\in}{T^\out}\frac{\omin}{\orin}-\frac{\omout}{\orout}\right)
\right]\;
\ee
and using the fact that along a particular worldline
\be
\eta=\frac{\pi^4}{30\zeta(3)}\left(\frac{T_0}{m_p}\right)\frac{\rho_b}{a\rho_\gamma}\propto \frac{T_0f_b\Omega_m}{\Omega_r},
\ee
we finally have
\bea
\delta_m&\simeq &\left(1-\frac{f_b^\out\eta^\in}{f_b^\in\eta^\out}\right)\left[1-\ain \frac{T_0^\in\omin}{T_0^\out\omout}\right],\\
\delta_r&\simeq &\frac{4}{3}\frac{\omout}{\orout}\left(1-\frac{f_b^\out\eta^\in}{f_b^\in\eta^\out}\right)\ain.
\eea
From this we see the results we present graphically, which are presented for the asymptotically flat case with $f_b=\,$constant: the radiation inhomogeneity is growing linearly with the scale factor, while the matter inhomogeneity is frozen very early,  decaying slightly from an initial value $\left(1-\frac{\eta^\in}{\eta^\out}\right)$. In the limit $\eta^\in=\eta^\out$, assumed in the standard model, all this behaviour disappears, and inhomogeneity grows at the next order in this series expansion; in particular, the matter and radiation inhomogeneities grow at the same rate. 

Let us discuss Fig.~\ref{fig:delta_evol} in more detail. Moving forwards in time, start with an under-density in the central matter compensated by a tiny over-density in the radiation, required for the models considered here. 
The matter density grows relative to radiation as we leave the radiation era, a process which starts to smooth out existing inhomogeneity in the matter, while decreasing the rate of growth of the inhomogeneity in the radiation. At the centre of the void the curvature starts to become significant at $t\sim 1$ Gyr at the expense of the matter component, but this doesn't happen so fast where the curvature starts smaller, at large distances such as at the CMB (where we observe $\eta^\out$).  In the language of perturbation theory, the curvature mode enters the Hubble radius and starts to grow. The growth of the curvature at the centre amplifies the late time inhomogeneity and is the process which accounts for the SNIa magnitudes.
The key finding is that \emph{the difference in the baryon density between the centre of the void and the outside is significant and $\mathcal{O}(1)$ at BBN time.} While the model homogenises as $t\to0$, it does so only in the dominant radiation component, and not in the matter. Only when the matter becomes relativistic will the model truly homogenise.

The $\mathcal{O}(1)$ difference in the baryon density inside and outside the void implies an $\mathcal{O}(1)$ difference in the spatial profile $\eta$ at BBN~--  exactly what we measure. Furthermore, SNIa observations imply a \emph{decrease} in the local matter density, the same as $^7$Li observations. 
Hence, a void-like profile for $\eta$ with our position close to the centre of the void can easily give an explanation to the $^7$Li problem if $\eta_{CMB}=\eta^\out\sim 1.5\times\eta^\in$; this model then evolves into a void which can explain dark energy.\footnote{A recent analysis~\cite{Holder:2009gd} constrained large-scale inhomogeneities in $\eta$ using D abundances (observed in different directions of the sky) and galaxy cluster gas fraction~\cite{Allen:2007ue} (observed up to $z\sim1$). Because we consider a spherically symmetric inhomogeneity and a homogeneous baryon fraction $f_b=\Omega_b/\Omega_m$ (with $f_b\sim0.17$ from the CMB), these bounds are automatically satisfied.} 
Relevant BBN processes depend only on the FLRW parameters for each patch; since we do not consider exotic contributions to Hubble rate or neutrino-antineutrino asymmetry, the deviation from standard BBN is only given by an inhomogeneous profile for $\eta$ giving different abundances in each patch.\footnote{Our inhomogeneous BBN is completely different from the inhomogeneous BBN relying on a varying baryon to photon and/or neutron to proton ratio on scales comparable to particle diffusion horizon at BBN~\cite{Lara:2006cd}. We also differ from non-standard BBN with large but not giant scale inhomogeneities (or anisotropies)~\cite{Malaney:1993ah} (i.e., scales smaller than (or comparable to) astrophysical structure scales), where primordial abundances produced in different regions are then mixed by structure formation.} The spatial distribution of nuclei primordial abundances today reflects the inhomogeneity at BBN time, as there is no homogenisation by structure formation on Gpc-scales. Note that we can fine-tune the initial conditions such that the lithium problem is solved, but without creating a void at late times (by enforcing a constant-curvature model). This requires an $\mathcal{O}(1)$ difference in the radiation density profile today, and a total density profile inhomogeneity which grows to $\sim 0.05$ at equality, then decays. This is much larger than a standard density perturbation at equality. Thus, there appears no way to solve the lithium problem within the standard FLRW models using this setup, though it might be possible within perturbed FLRW models with a large spherical isocurvature mode.

\section{An example}

In the previous Section, we described how models with Hubble-scale inhomogeneity can solve the $^7$Li problem.
Now we show with an example that this is not only a general possibility, but rather the actual parameters involved in this scenario are in the same ball-park required by other cosmological observations, such as Hubble rate measurements, supernovae observations and the CMB. The calculation presented here is discussed in detail in~\cite{cmbpaper}.

Voids may be described by a profile for the present-day matter density $\Omega_m(r)=\omout-(\omout-\omin)\,e^{-r^2/2\sigma^2}=1-\Omega_k(r)$, and a similar profile for the radiation density (usually ignored but important here).  The Hubble rate $H(t,r)$ is given by a generalized Friedman equation
(see Appendix in~\cite{cmbpaper}) 
%$H(t,r)=\partial\ln a(t,r)/\partial\tau=100 h(r)\sqrt{\Omega_r a(t,r)^{-4}+\Omega_m a(t,r)^{-3}+\Omega_k a(t,r)^{-2}}$ km s$^{-1}$Mpc$^{-1}$, where $\tau$ is the proper time along a worldline 
and we choose $a(t_0,r)=1$ to fix the gauge freedom in void models. 
For our purposes, we can calculate nearly everything along the central worldline and along one at the radius of the CMB ($r\gg\sigma$), neglecting all $r$-dependence, and use the standard Friedman equation. The exception is the area distance to the CMB, which is calculated in the normal way (see e.g.,~\cite{February:2009pv}), and is affected by the shape and width of the void. 
Indeed CMB fluctuations at small scales (and focusing on the first peaks) are mainly set by three parameters~\cite{Hu:2001bc,Wang:2007mza,Komatsu:2008hk}: the angular scale of the sound horizon $l_a$ , the particle horizon at matter-radiation equality $l_{eq}$, and the baryon-photon density ratio at decoupling $\displaystyle R_*=\left.{3\,\rho_b}/{4\,\rho_{\gamma}}\right|_{\dec}$, where $l_a/l_{eq}$ and $R_*$ depend only on local physics at decoupling, and then the area distance to the last scattering surface enters in the same way in $l_a$ and $l_{eq}$ (for a detailed analysis, see~\cite{cmbpaper}).
In the example we are going to present, we enforce $l_a=302$, $l_{eq}=136$, and $R_*=0.63$, in near perfect agreement with WMAP data~\cite{WMAP5par}.  

We fix the void parameters on the inside and outside using sample observations as follows:

%CMB fluctuations in the first peaks are mainly set by three parameters~\cite{Hu:2001bc,Wang:2007mza,Komatsu:2008hk}: the angular scale of the sound horizon $l_a=\pi\, d_A(z_{\dec})/[a_{\dec} r_s(a_{\dec})]$ ($r_s$ the sound horizon at photon decoupling with scale factor $a_{\dec}$, and $d_A(z_{\dec})$ is the area distance to the last scattering surface), the particle horizon at matter-radiation equality $l_{eq}=k_{eq}d_A(z_{\dec})/a_{\dec}$, and the baryon-photon density ratio at decoupling $\displaystyle R_*=\left.{3\,\rho_b}/{4\,\rho_{\gamma}}\right|_{\dec}$. 

%\begin{itemize}

\noindent {\bf Local constraints:} Take $T_0^\in\approx2.725K\Rightarrow \Omega_\gamma^\in h_\in^2\approx 2.49\times10^{-5}$ from the CMB temperature today (which gives $\Omega_r^\in\approx1.69\Omega_\gamma^\in$ for $N_{\mathrm{eff}}=3.04$); $\eta^\in\sim4.5\times 10^{-10}$ from $^7$Li measurements; $h^\in=0.74\pm4$ from HST measurements~\cite{Riess:2009pu}. We expect $\omin\sim 0.15$ from SNIa constraints.  The age is $t_0=\int_0^1\frac{da}{aH}$ along the central worldline. We assume the Bang time to be homogeneous $t_B=0$, so $t_0$ is the age of the universe everywhere today.
%\footnote{There is no particular reason to restrict this degree of freedom in inhomogeneous models, and having $t_B(r)\neq0$}

\noindent {\bf Asymptotic constraints:} Given CMB observations of $l_a, l_{eq}$ and $R_*$, we may fix $\eta^\out$ and $f_b^\out$ from $R_*$ and $l_a/l_{eq}$, and then find $d_A(z_{\dec})$ from $l_a$ or $l_{eq}$. 
This implies $\eta^\out\approx6.2\times10^{-10}$, and $f_b^\out\approx0.17$~\cite{cmbpaper} (in our example, we assume $f_{b}=$const., so that $f_{b}^\in=f_{b}^\out$).
We can also require $\eta^\out\approx6.2\times10^{-10}$ to be compatible with D observations at $z\sim3$ (although the void profile and width can affect this, if the asymptotic value is not reached by this redshift). The redshift of decoupling is defined by $1+z_{\dec}=T_{\dec}^\out/T_0^\in$, where the local temperature of the CMB, $T_{\dec}$, can be in first approximation determined when the Thompson scattering interaction rate $\Gamma=H^\out$, and making use of the Saha equation, or more precisely through a fitting formula~\cite{cmbpaper}.

\noindent {\bf  Constraints from distances:}
The matter profile is constrained by  measurements of distances.
The matter density has to be $\Omega_m=1-\Omega_k\sim0.1-0.2$ inside the void at the centre, which must raise to $\Omega_m\sim0.5$ by $z\sim1$, or a (co-moving) distance of a few Gpc, to satisfy SNIa observations~\cite{February:2009pv,Sollerman:2009yu} (then may or may not reach asymptotic flatness).
Our void must then be able to fit the area distance to the last scattering surface $d_A(z_{\dec})\approx13\,$Mpc derived from CMB observations, which constrains $\sigma$ and $\Omega_m^\out$. 
Consistent pictures typically require $\sigma\sim$ few Gpc, and asymptotically open models, although considering profiles more complicated than a Gaussian (or a significantly lower $h^\in$), $\Omega_m^\out\gtrsim1$ is also possible~\cite{cmbpaper}.

\begin{figure*}[t]
\begin{center}
\includegraphics[width=1\textwidth]{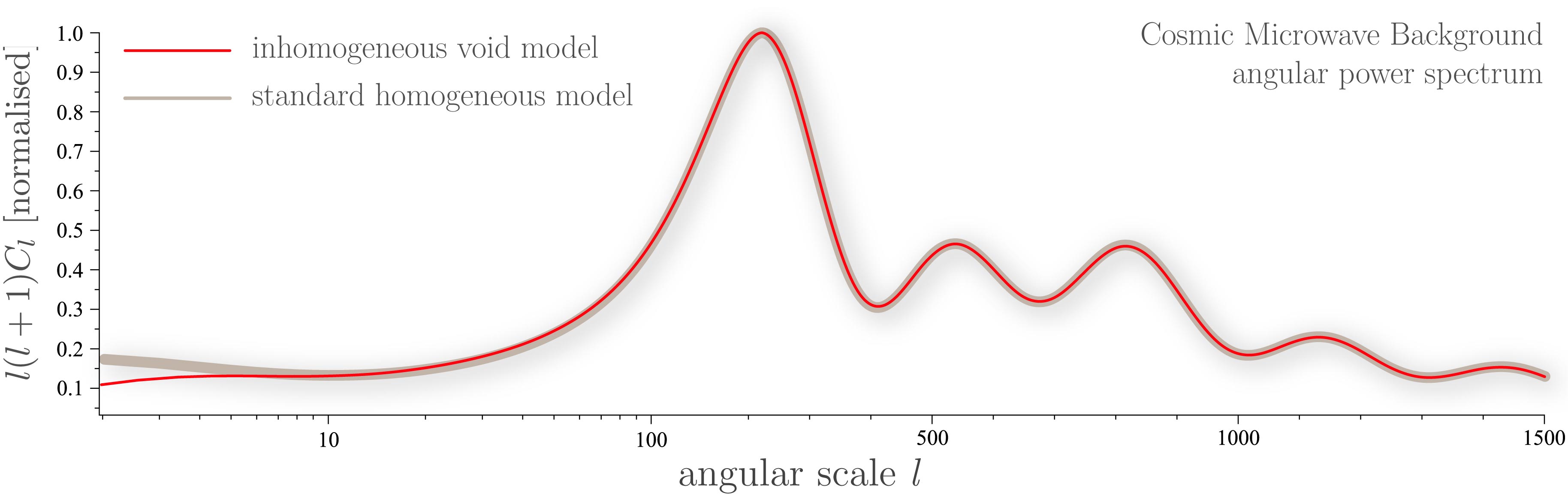}
\caption{{\bf The normalised CMB angular power spectrum.} This is for the example model described in the text, shown on a log scale for $l<200$ and linear for $l>200$.  The power spectrum is shown against a default flat concordance model with zero tilt, and $h=0.7, \Omega_m h^2=0.137, \Omega_b h^2=0.0226$. There's nothing between the two models for high $l$, with the maximum difference around 1\%. This is also the case for the polarisation spectra (not shown). }
\label{cmb}
\end{center}
\end{figure*}

\noindent {\bf Derived parameters:} 
The baryon density is given by:
$
\Omega_b^\in h_\in^2\approx 3.70\times10^6\eta^\in({T_0^\in}/{2.7\text{K}})^3.
$
This fixes $f_{b}^\in=\Omega_b^\in/\omin$. We may estimate $h^\out$ from $\displaystyle t_0=\int_0^1\frac{d\aout}{\aout H^\out}$, ignoring the contribution from $\orout$ which we don't yet know. This introduces a tiny error, which can be accounted for iteratively. The effective FLRW temperature of the CMB today at the radius where we see the CMB emitted is:
\be
\frac{T_0^\out}{T_0^\in}=\left(\frac{\Omega_m^\out h_\out^2}{\Omega_m^\in h_\in^2}\frac{\eta^\in}{\eta^\out} \right)^{1/3}.
\ee
Note that $T_0^\out$ is significantly different from $T_0^\in$.
Then $\Omega_\gamma^\out h_\out^2\approx 4.48\times10^{-7}(T_0^\out/1\text{K})^4$ follows. All `out' parameters which are set at $t=t_0$ refer to an asymptotic observer (who can also be seen as defining an effective FLRW model) which has the same intrinsic CMB (though usually different area distance) as the void observer will see~\footnote{Mostly, however, they will be close to the real values along the wordline of the CMB (but note that they aren't actually required in this analysis). For example, a real outside observer (i.e., along the wordline of the CMB) will see a cold spot in the CMB where the void is, which will lower the mean `outside' temperature. }. To calculate the full CMB spectrum, calculate the shift $S=d_A^{LTB}(z=T_{\dec}^\out/T_0^\in)/d_A^{FLRW\out}(z=T_{\dec}^\out/T_0^\out)$ for the angular power spectrum output from a standard CMB code set to the `out' values (importantly using $T_0^\out$) to match the data~\cite{Clifton:2009kx}. This gives $C_\ell^{\text{void}}\propto C_{\ell/S}^{{FLRW\out}}$, with any amplitude shift being absorbed by the unknown amplitude in the primordial power spectrum. The result is shown in Fig.~\ref{cmb}.

%\end{itemize}

\begin{figure}[t]
\begin{center}
\includegraphics[width=0.7\columnwidth]{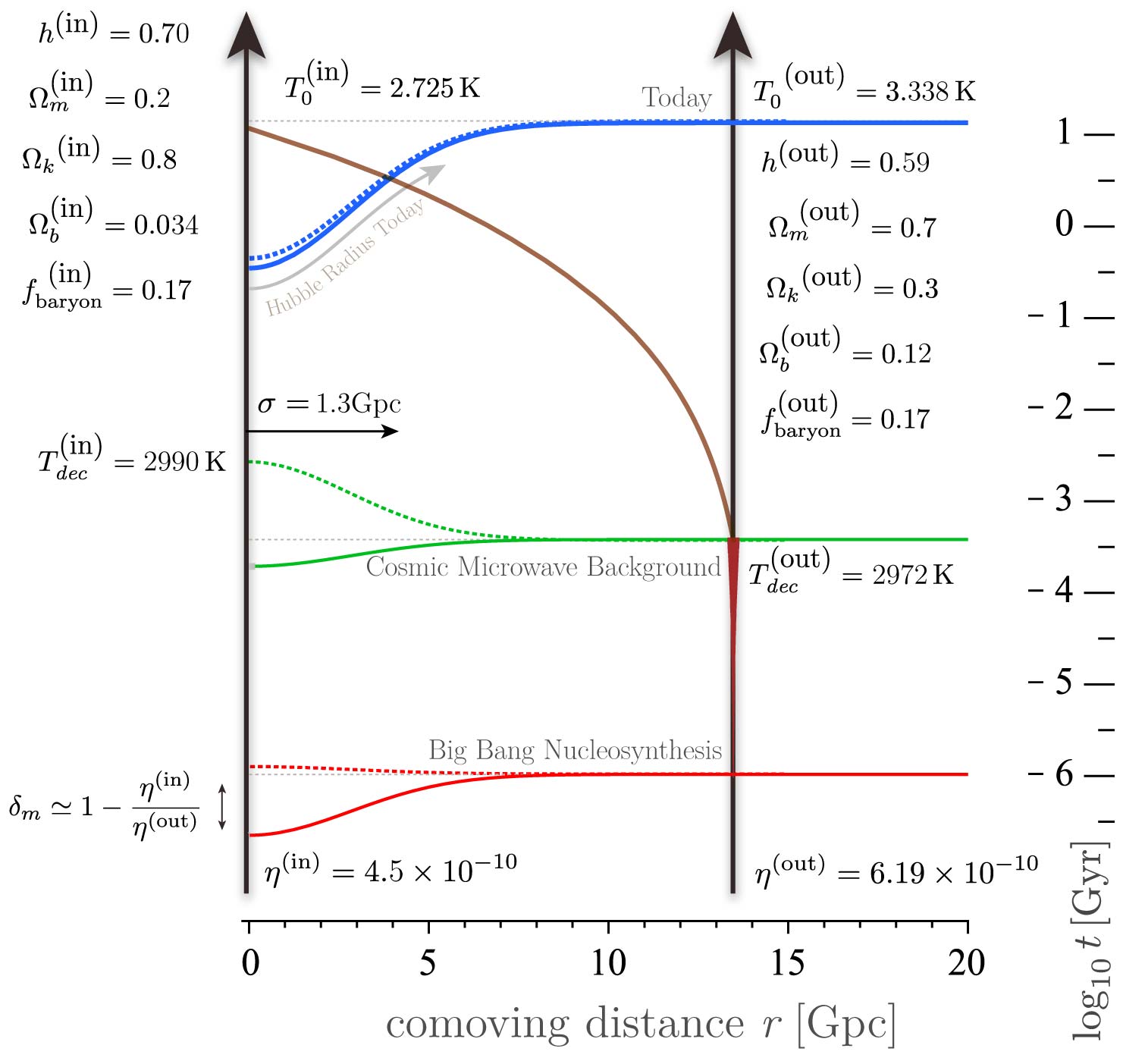}
\caption{{\bf A spacetime overview of an example model.} At the three key times we show the matter profile (solid). The radiation profile (dashed) at early times is an over density in this scenario.}
\label{overview}
\end{center}
\end{figure}

From this procedure one can construct a void model in agreement with CMB, SNIa, HST and BBN observations (see Fig.~\ref{overview})~\footnote{Age constraints~\cite{ages} favour a slightly lower value of $h$ than the HST value, but with $\Omega_m$ low enough to satisfy SNIa data, these are relatively easy to satisfy.}. 
As an example, we choose $h=0.70$,  $\eta^\in=4.5\times10^{-10}$, $f_b=$const. (which implies $\Omega_m^\in\approx0.2$) a Gaussian profile which has a full width at half maximum  of $\sim3\,$Gpc and has $\Omega_m\approx0.7$ at the CMB distance. It provides an excellent fit to the CMB, as shown by the angular power spectrum in Fig.~\ref{cmb}. It is interesting to note that if we make $\eta^\in$ larger than this, then it makes $\Omega_m^\in$ too high for the SNIa (we loose the `bump' in the distance modulus)~-- so, in effect the SNIa favour a low $\eta^\in$. \footnote{Previous analysis of the CMB in void models found that the CMB couldn't be accounted for except in very unusual configurations~\cite{Zibin:2008vk,Clifton:2009kx}, although~\cite{Clifton:2009kx} show that an inhomogeneous bang time allowed a wide class of voids to fit. Their analysis is based on test radiation in pure dust void models, and we find that including inhomogeneous radiation in the model allows a much wider class of voids to fit the CMB~\cite{cmbpaper}.}  
For a more detailed description of our CMB computation see Ref.~\cite{cmbpaper}.

This model fits the $^7$Li constraints in the most natural way: it requires an $\mathcal{O}(1)$ decrease in the baryon density at the centre, just as dark energy requires an $\mathcal{O}(1)$ decrease in the CDM. While there still exists freedom in the models to fit more complicated possibilities (such as an inhomogeneous bang time or $f_b$)\footnote{Note that they can also be invoked in order to compensate for possible deviations from FLRW evolutions we consider at the center and asymptotically.}, we do not need to invoke them. (A full likelihood analysis will be explored elsewhere.) There is no fine-tuning involved in these fits. By contrast, the assumption in the standard model that $\eta$ is constant across comoving volumes $\sim\,$Gpc$^3$ may be considered extremely fine-tuned in an inhomogeneous context. 

The large scale CMB together with the Baryon Acoustic Oscillations (BAO) depend on the detailed evolution of perturbations during the curvature era which are not yet understood in the context of inhomogeneous models. Moreover to extract constraints from BAO and kinetic Sunyaev-Zeldovich a detailed treatment of the radiation is necessary~\cite{cmbpaper}.

While important for CMB-related measurements, the evolution of radiation at late times is completely irrelevant for our BBN analysis. For example, suppose one can construct models where the temperature of the radiation does not evolve as $1/a$ along the central worldline at late times (i.e., in the curvature era), which leads to non-conservation of $\eta$ (evaluating the radiation temperature using the geodesics of the LTB solution leads to this conclusion). While in this case the CMB predictions will be different~\cite{Zibin:2008vk,Clifton:2009kx},
%and necessitates the introduction of the bang time function~\cite{Clifton:2009kx}, 
the core of our analysis is unchanged because $^7$Li abundances constrain $\eta^\in$ at early times, when FLRW evolution in each causal patch is certainly a good approximation for a Gpc-scale inhomogeneity. In particular, Fig.~\ref{fig:delta_evol} and surrounding analysis is unchanged except in the curvature era. 

\section{Conclusions}
We have found that the inhomogeneous models which explain dark energy without exotic physics at late times also explain the lithium abundances in a natural way.  In retrospect, the fact that $^7$Li gives us an unusual glimpse of our past worldline implies that this is actually a test of homogeneity at very early times. We have argued that, while there may exist an explanation of the lithium abundances in terms of an anomalously large Gpc isocurvature mode in FLRW, the fact that this must be spherical or otherwise localised to avoid problems with the CMB implies a strong link with the spherical inhomogeneous models for dark energy. The origin of the inhomogeneous profile lies in the primordial baryon asymmetry-entropy density ratio, and may in fact be laid down at the onset of an inflationary period which lasts slightly less than the 60-efolds or so required for the standard model; it cannot plausibly exist within standard slow roll inflation.  At first sight the void models appear fine-tuned as we must be within tens of Mpc of the centre~\cite{alnes,mortsell}, a coincidence of 1 part in $\sim10^{8}$. 
But this has to be compared to coincidences of $10^{120}$ or even $10^{500}$ encountered in the standard model. In any case, the spherically symmetric void models are merely a simple first attempt at an inhomogeneous universe and so should be understood as toy models in this context. We would argue that more sophisticated models of inhomogeneity should be explored to address the Copernican problem, and that the problems of structure formation and inflation should now be properly investigated.

\acknowledgments

We thank Bruce Bassett, Timothy Clifton, George Ellis, Pedro Ferreira, Roy Maartens and Jean-Philippe Uzan for discussions. This work is supported by the National Research Foundation (South Africa). M.R. acknowledges funding by the Centre for High Performance Computing, Cape Town and the hospitality of the African Institute for Mathematical Sciences.

\section*{Note added}

Since the first version of this paper appeared on the arXiv, a number of other papers have appeared which also discuss the CMB in void models~\cite{Yoo1,BNV,MZS,Yoo2} (see also~\cite{Vonlanthen}). Comparisons with our work are discussed in our follow-up paper~\cite{cmbpaper}.

%%%%%%%%%%%%%%%%%%%%%%%%%%%%%%%%%%%%%%%%%%%%%%%%%%%%%%%%%%%%%%%%%%%%%%%%%%%%%%%

\end{document}